\documentclass[aps,prb,citeautoscript,preprint,noshowpacs,superscriptaddress]{revtex4-1}  

\usepackage{graphpap}
\usepackage[dvips]{graphicx}
\usepackage[dvips]{graphics}
\usepackage{color}
\usepackage{hyperref}
\usepackage{amsmath}
\usepackage[normalem]{ulem}

\begin{document}

\title{Optomechanical Measurement of Thermal Transport in Two-Dimensional MoSe$_2$ Lattices}
\author{Nicolas Morell}
\author{Slaven Tepsic}
\author{Antoine Reserbat-Plantey}
\affiliation{ICFO - Institut de Ciencies Fotoniques, The Barcelona Institute of Science and Technology, 08860 Castelldefels, Barcelona,Spain}
\author{Andrea Cepellotti}
\affiliation{Department of Physics, University of California at Berkeley and Materials Sciences Division, Lawrence Berkeley National Laboratory, Berkeley, California 94720, United States}
\author{Marco Manca}
\affiliation{Universit\'e de Toulouse, INSA-CNRS-UPS, LPCNO, 135 Avenue Rangueil, 31077 Toulouse, France}
\author{Itai Epstein}
\affiliation{ICFO - Institut de Ciencies Fotoniques, The Barcelona Institute of Science and Technology, 08860 Castelldefels, Barcelona,Spain}
\author{Andreas Isacsson}
\affiliation{Department of Physics, Chalmers University of Technology, S-41296 G\"oteborg, Sweden}
\author{Xavier Marie}
\affiliation{Universit\'e de Toulouse, INSA-CNRS-UPS, LPCNO, 135 Avenue Rangueil, 31077 Toulouse, France}
\author{Francesco Mauri}
\affiliation{Dipartimento di Fisica, Universit\`a di Roma La Sapienza, Piazzale Aldo Moro 5, I-00185 Rome, Italy}
\author{Adrian Bachtold}
\affiliation{ICFO - Institut de Ciencies Fotoniques, The Barcelona Institute of Science and Technology, 08860 Castelldefels, Barcelona,Spain}

\begin{abstract}

Nanomechanical resonators have emerged as sensors with exceptional sensitivities. These sensing capabilities open new possibilities in the studies of the thermodynamic properties in condensed matter. Here, we use mechanical sensing as a novel approach to measure the thermal properties of low-dimensional materials. We measure the temperature dependence of both the thermal conductivity and the specific heat capacity of a transition metal dichalcogenide (TMD) monolayer down to cryogenic temperature, something that has not been achieved thus far with a single nanoscale object. These measurements show how heat is transported by phonons in two-dimensional systems. Both the thermal conductivity and the specific heat capacity measurements are consistent with predictions based on first-principles.

\end{abstract}
\maketitle

Mechanical resonators based on suspended nanoscale objects, such as monolayer semiconductors~\cite{lee2013,Castellanos2013,Morell2016,Lee2018}, graphene~\cite{Bunch2007,Chen2009,eichler2011a,Miao2014,Singh2014,Song2014,Weber2014,Cole2015,DeAlba2016,Mathew2016,Guttinger2017,Will2017}, nanotubes~\cite{Sazonova2004,Lassagne2009,Steele2009,Gouttenoire2010,Chaste2012,Stapfner2013,Moser2013,Ganzhorn2013,Benyamini2014,Tavernarakis2018,Bonis2018,Khivrich2019}, and semiconducting nanowires~\cite{Ayari2007,Gil-santos2010,Nichol2013,Gloppe2014,Nigues2015,Sansa2016,Rossi2016,Lepinay2016}, have attracted considerable interest.
Because of their small mass, such resonators become fantastic sensors of external forces and the adsorption of mass~\cite{Hanay2012,Chaste2012,Moser2013,Bonis2018}. The sensing capabilities of nano- and micro-resonators have been used with great success in recent advances of various fields. These include nano-magnetism~\cite{Losby2015,Mehlin2015}, surface imaging~\cite{Rossi2016,Lepinay2016}, surface science~\cite{Wang2010,Tavernarakis2014}, light-matter interaction~\cite{Gloppe2014},  persistent currents in normal metal rings~\cite{Bleszynski-Jayich2009}, and engineered electron-phonon coupling~\cite{Benyamini2014}. In this work, we show how optomechanical systems can be used to study heat transport in individual low-dimensional materials.

Heat transport at the nanoscale is of major fundamental interest for a broad range of research fields, such as nanophononics~\cite{Schwab2000,Li2012,Tavakoli2018}, spintronics~\cite{Bauer2012}, quantum electron devices~\cite{Giazotto2006,Jezouin2013}, and quantum thermodynamics~\cite{Pekola2015}. Heat can be controlled and measured with good accuracy in devices micro-fabricated from bulk material. By contrast, heat transport in devices based on low-dimensional materials cooled at low temperature is still at its infancy. Measuring their thermal conductance at cryogenic temperature is a challenging task. It requires the fabrication of sophisticated devices, which incorporate local heaters and thermometers, and a careful calibration of the latter~\cite{Xu2014,Kim2001}. The difficulty of fabricating reliable devices has hindered progress in the field for many years.

Lattice vibrations  are the main carriers of heat in a large variety of low-dimensional materials, including carbon nanotubes~\cite{Kim2001,Yu2005}, graphene~\cite{Balandin2011}, and semiconductor monolayers~\cite{Yan2014}. Heat transport has been intensively studied at room temperature and above using Raman measurements~\cite{Balandin2008a,Cai2010,Ghosh2010,Faugeras2010,Lee2011,Zhang2015,Sledzinska2016,Ye2018} and scanning probe microscopy~\cite{Shi2009,Pumarol2012}. Heat transport enters into interesting regimes at low temperature, such as the dissipationless transport through low-dimensional materials in the ballistic regime~\cite{Xu2014,Kim2001,Bae2013} and the phonon hydrodynamic regime predicted in monolayers~\cite{Cepellotti2015,Lee2015b}. The interpretation of heat transport measurements can be difficult, since the thermal conductance depends on various quantities that have not been measured independently thus far. These include the heat capacity and the phononic mean-free path. Recently, new methods have been reported to measure the electron contribution of the thermal conductivity of graphene down to low temperature~\cite{Fong2016,Efetov2018}.

Heat transport measurements in low-dimensional materials have mainly consisted in probing the thermal conductance $K$, that is, how well the system conducts heat. In optomechanics, it is possible to measure how quickly the mechanical resonator conducts heat~\cite{Metzger2008,Schwarz2016,Dolleman2017}. The characteristic time $\tau$ for the heat to travel out of the resonator introduces a retarded force acting on the mechanical resonator~\cite{Metzger2004,Barton2012}.

Here, we combine two methods to measure $K$ and $\tau$ in a optomechanical resonator based on a vibrating MoSe$_2$ monolayer. This allows us to unravel the thermal properties of low-dimensional materials down to cryogenic temperature and with a device that is simple to fabricate. Our measurements indicate that the phonon transport is diffusive above $\sim 100$~K, while the majority of phonon carriers are ballistic over the size of the device at low temperature. The temperature dependence of the specific heat capacity approaches a quadratic dependence, the signature of two-dimensional lattices. Both the thermal conductance and the specific heat capacity measurements can be described by predictions based on first-principles.

The mechanical resonator consists of a MoSe$_2$ monolayer drum (Fig. 1a-d). The drum is fabricated with the dry transfer of MoSe$_2$ monolayers using a polydimethylsiloxane (PDMS) stamp~\cite{Castellanos2014} over a highly doped Si substrate with prestructured holes. MoSe$_2$ monolayers are obtained from mechanical exfoliation of crystals purchased from 2D Semiconductors. The device is measured in a cryostat whose temperature can be set between 3 and 300~K. Photoluminescence spectra at 3~K feature narrow peaks associated with two-dimensional excitons and trions with a wavelength at $\sim 762$~nm and $\sim 748$~nm, respectively (Section 1 of Supplementary Information), in agreement with previous reports~\cite{Ross2013,Wang2015}. Photoluminescence maps are homogeneous~\cite{Morell2016}. These measurements confirm that the drums are made from MoSe$_2$ monolayers. The metal electrode attached to the MoSe$_2$ flake is used to apply an electrostatic force on the drum (Fig.~1a,b); it has no effect on the thermal transport.

\begin{figure*}
	\includegraphics[width=17cm]{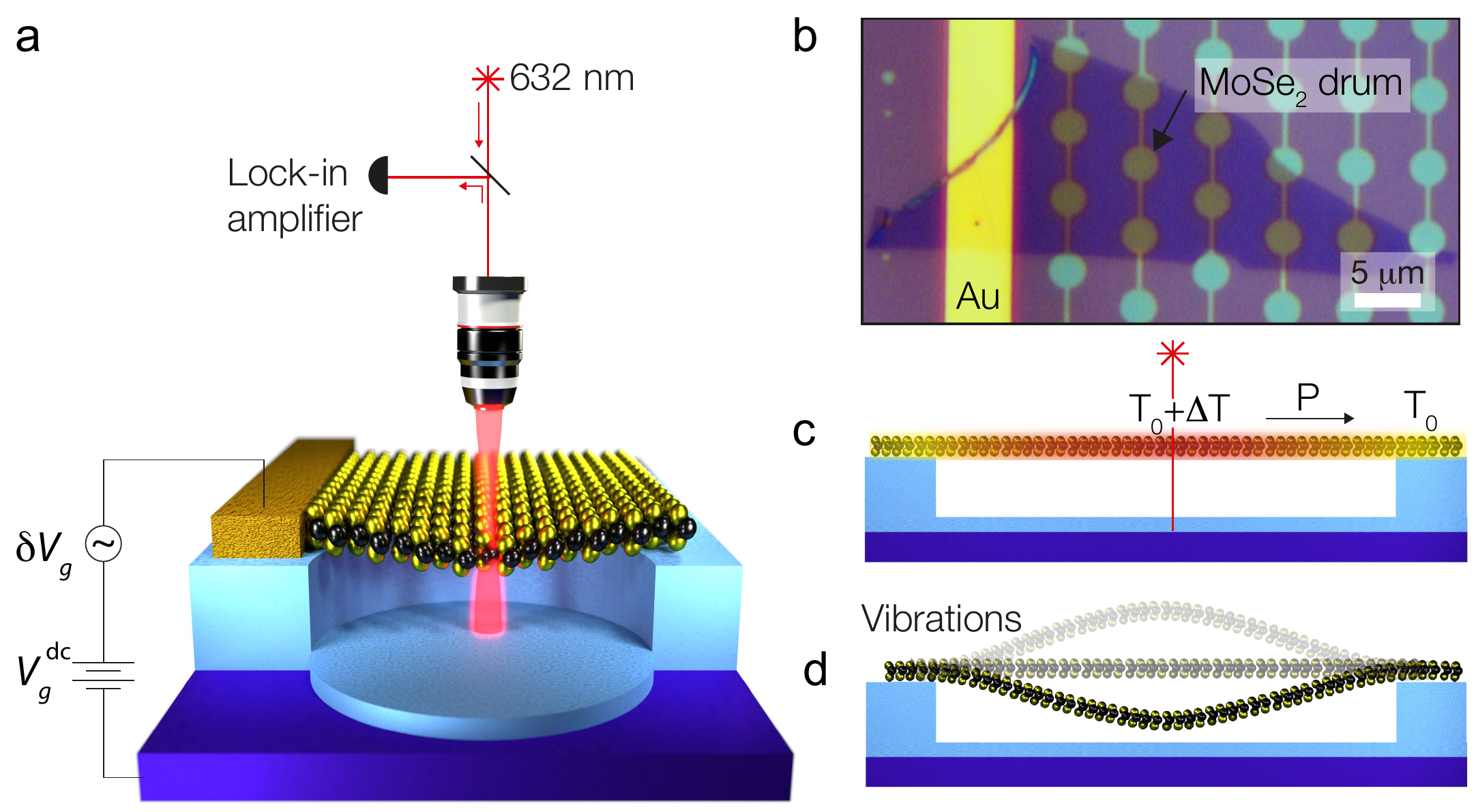}
	\caption{\textbf{Optomechanical device structure and working principle used to measure thermal transport.} (\textbf{a}) Schematic of the optomechanical device. The mechanical vibrations are driven capacitively and detected by optical interferometry~\cite{Morell2016}. The MoSe$_2$ monolayer is a mobile absorber in an optical standing wave produced by a 632~nm probe laser. The modulated laser reflection intensity is measured with an avalanche photo-detector feeding a lock-in amplifier. (\textbf{b}) Optical microscopy image of a typical device. (\textbf{c}) Heat transport induced by the absorption of the laser power. A temperature difference $\Delta T$ is created from the heat flow. (\textbf{d}) Detection of the laser-induced temperature rise $\Delta T$ using the fundamental mechanical mode of the optomechanical resonator.}
	\label{fig1}
\end{figure*}

Mechanical vibrations are detected by optical interferometry~\cite{Barton2012,Morell2016}. A continuous wave laser impinges on the center of the MoSe$_2$ membrane, and the reflected laser light intensity is modulated by an amount proportional to displacement of the resonator. The laser forms a standing wave pattern in the direction perpendicular to the Si substrate, such that the displacement of the monolayer modifies its optical absorption. The laser spot has a measured radius of about 350~nm. The fact of measuring the mechanical resonator with the laser beam modifies the dynamics of the mechanical vibrations by a small amount -- increasing the laser power modifies the resonance frequency and the resonance linewidth (Fig. 2). This backaction has two components, the static and the dynamical backaction. The former allows us to quantify $K$, and the latter $\tau$.

\begin{figure*}
	\includegraphics[width=12cm]{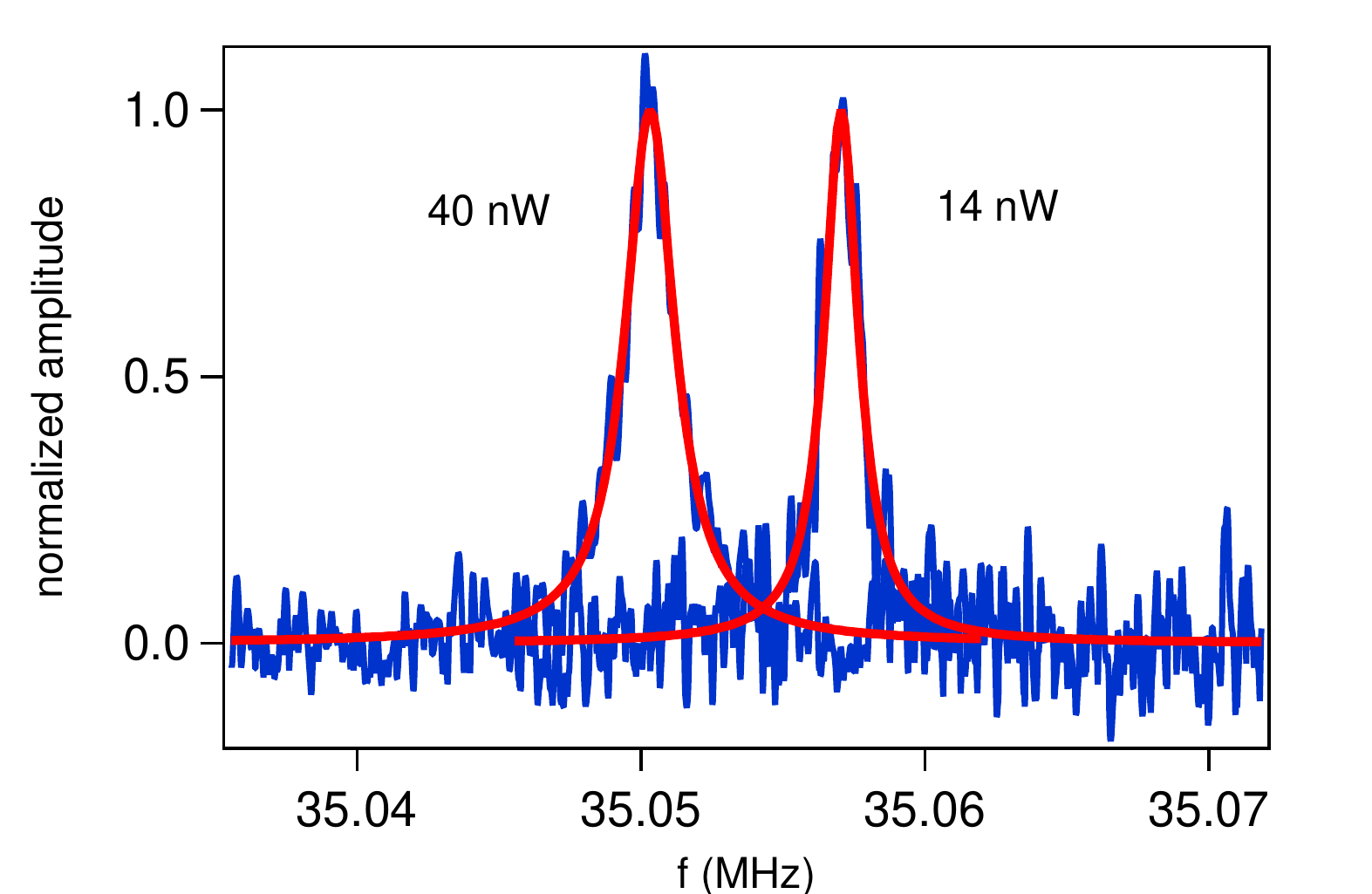}
	\caption{\textbf{Backaction of the laser on mechanical vibrations.}  Response of the displacement amplitude of the mechanical mode as a function of the frequency of the driving force for two different absorbed laser powers. The temperature is set at 3~K and the gate voltage at 4~V. The red lines correspond to Lorentzian fits.}
	\label{fig2}
\end{figure*}

We measure the thermal conductance in a way similar to the well-established method based on Raman measurements employed at room temperature~\cite{Balandin2008a,Cai2010}. The static backaction of the laser beam is a simple absorption heating effect, which results in a temperature gradient $\Delta T$ between the center of the membrane and its circular clamp (Fig. 1c). The heat flow is given by the power $P$ absorbed in the membrane (Section 2 of Supplementary Information). In a Raman measurement, $\Delta T$ is quantified by the frequency shift of Raman-active peaks. In our case, $\Delta T$ is measured by the frequency shift $\Delta f_\mathrm{T}$ of the fundamental mechanical mode (Fig. 1d). As a result, the equivalent thermal conductance is $K=P/\Delta T$. Mechanical MoSe$_2$ drums with their high quality factor~\cite{Morell2016} are extremely good temperature sensors, allowing us to measure the linear thermal conductance down to 3~K. This is a significant improvement compared to Raman measurements, which are typically operated at 300~K or above, because the detection of the frequency shift of Raman-active peaks requires comparatively large $P$.

We measure $\tau$ from the effect of the dynamical backaction on the electrostically driven vibrations. Absorption heating from the laser beam expands the MoSe$_2$ crystal~\cite{Morell2016}, which is equivalent to a force acting on the membrane. The crystal expansion responds to a change in the absorbed laser power with delay, that is, the time $\tau$ for the membrane to heat up or to cool down. The absorbed laser power oscillates in time because of the oscillating motion of the membrane in the laser interference pattern used to detect the vibrations. Overall, the photothermal force oscillates with a finite phase shift compared to the motion of the membrane. The in-phase photothermal force modifies the resonance frequency by $\Delta f_\mathrm{B}$ and the out-of-phase photothermal force modifies the resonance linewidth by $\Delta \Gamma_\mathrm{B}$ as
\begin{eqnarray}
&& \Delta f_\mathrm{B}=-\frac{1}{2} f_\mathrm{m} \frac{\partial_\mathrm{z}F^\mathrm{z}_\mathrm{photo}}{k}\frac{1}{1+(2\pi f_\mathrm{m}\tau)^2},
\label{DfB}\\
&& \Delta \Gamma_\mathrm{B}=f_\mathrm{m} \frac{\partial_\mathrm{z}F^\mathrm{z}_\mathrm{photo}}{k}\frac{2\pi f_\mathrm{m}\tau}{1+(2\pi f_\mathrm{m}\tau)^2}.
\label{GFB}
\end{eqnarray}
Here, $f_\mathrm{m}$ is the resonance frequency of the mechanical mode, $k$ its spring constant, $z$ the coordinate in the direction perpendicular to the membrane, and $\partial_\mathrm{z}F^\mathrm{z}_\mathrm{photo}$ the derivative of the $z$-component of the photothermal force with respect to $z$. We infer $\tau$ from $\Delta f_\mathrm{B}$ and $\Delta \Gamma_\mathrm{B}$ for a fixed laser power using $\tau= -\Delta \Gamma_\mathrm{B}/4\pi f_\mathrm{m}\Delta f_\mathrm{B}$.

The key to quantify $\Delta f_\mathrm{T}$ and $\Delta f_\mathrm{B}$ is to deform the static profile of the drum with an electrostatic force (Fig. 3a). The drum is straight when it is not subject to a sizeable electrostatic force. This is because the drum is mechanically stretched by the circular clamp, as shown by the strong temperature dependence of $f_\mathrm{m}$ (Figs. 3b,c); the tensile strain in the membrane is quantified by the measured dependence of $f_\mathrm{m}$ on the electrostatic force (Section 3 of Supplementary Information)~\cite{Morell2016}. The absorbed laser power generates a photothermal force $F_\mathrm{photo}$ that reduces the stretching force. When the drum is straight, the photothermal force is perpendicular to the motion of the vibrations, so that $\Delta f_\mathrm{B}=0$ (Eq.~\ref{DfB}); in this straight configuration, we only measure $\Delta f_\mathrm{T}$ associated to the thermal conductance. When the drum is bent, the photothermal force modifies both $\Delta f_\mathrm{T}$ and $\Delta f_\mathrm{B}$. We obtain $\Delta f_\mathrm{B}$ by subtracting the frequency shifts measured in the bending and the straight configurations (Section 4 of Supplementary Information). We go from a straight configuration to a bending configuration by applying a voltage $V_\mathrm{g}^\mathrm{dc}$ onto the gate electrode.

\begin{figure*}
	\includegraphics[width=12cm]{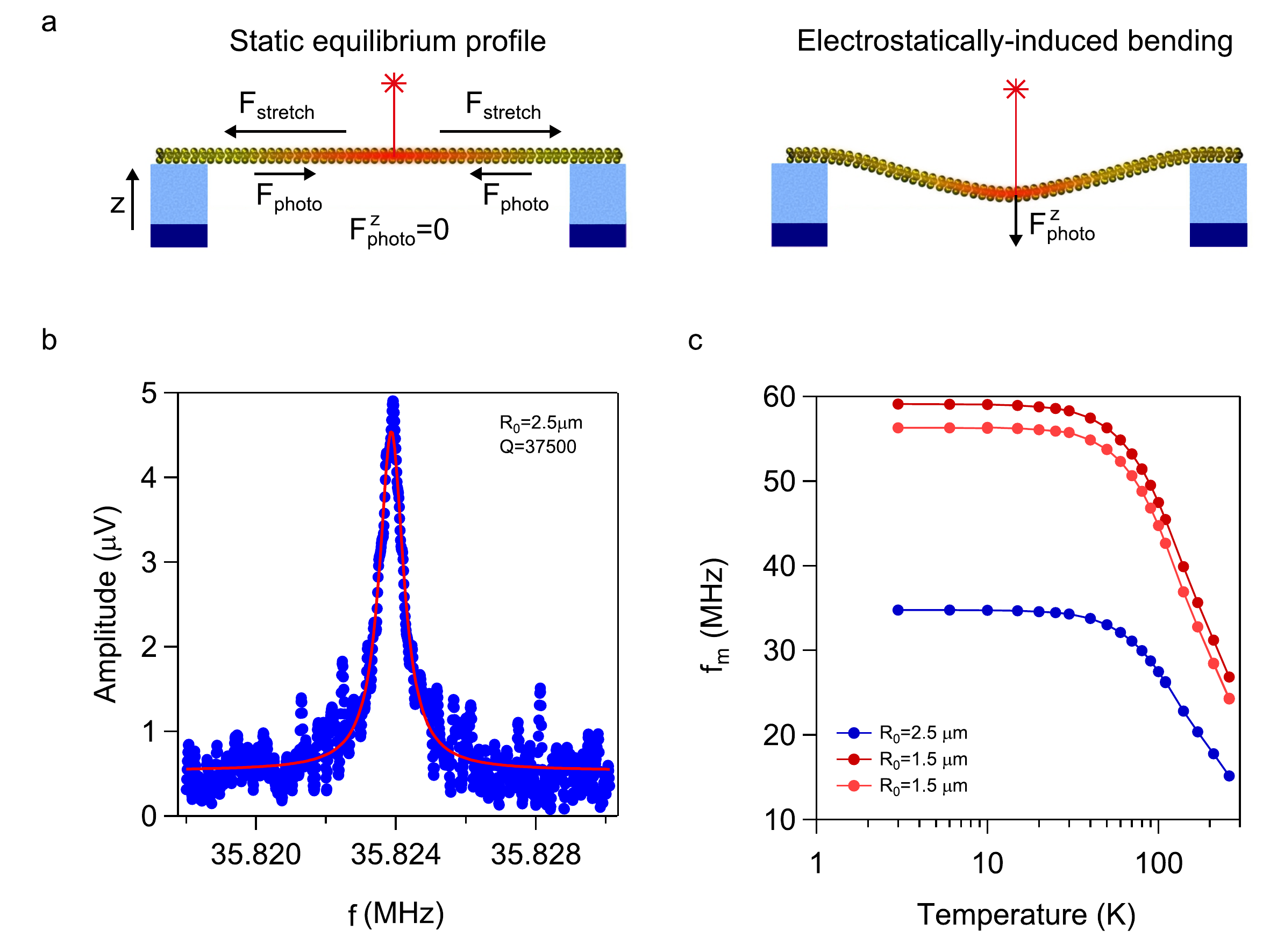}
	\caption{\textbf{MoSe$_2$ drum under mechanical tension.} (\textbf{a}) Static profile of the drum controlled with an electrostatic force, by applying the tension $V_\mathrm{g}^\mathrm{dc}$ on the backgate (Fig.~1a). In the straight configuration, the measured frequency shift is solely related to static backaction, which allows us to quantify the thermal conductance. The drum is stretched by the force $F_\mathrm{strech}$ from the circular clamp. The force $F_\mathrm{photo}$ produced by the laser beam reduces the stretching. In the bending configuration, the frequency shift also depends on dynamical backaction, because $\partial_\mathrm{z} F_\mathrm{photo}^\mathrm{z} $ is finite. This allows us to quantify the time $\tau$ for the heat to travel out from the drum. The amplitude of the mechanical vibrations ($<1$~nm) is smaller than the static displacement ($\lesssim 10$~nm) in the bending configuration. (\textbf{b}) Response of the displacement amplitude of the mechanical mode as a function of the frequency of the driving force at 3~K. (\textbf{c}) Resonance frequency of the mechanical mode as a function of temperature for three different devices when the drum is in the straight configuration.}
	\label{fig3}
\end{figure*}

Figures~4a,b show the temperature dependance of the equivalent thermal conductance. The conductance is obtained from the slope $\Delta f_\mathrm{m}/\Delta P$ in Fig.~4a using the calibration slope $\Delta f_\mathrm{m}/\Delta T$ in Fig.~3c. The conductance is measured in the linear regime, because the applied $P$ is low. The largest $\Delta T$ remains below 1~K. The estimation of the absorbed laser power is detailed in Section 2 of Supplementary Information; we use 5.7 \% for the absorption coefficient of MoSe$_2$ monolayers~\cite{Zhang2015}. We also show that the absorption coefficient is independent of temperature and gate voltage at the laser wavelength $\lambda=632$~nm. In order to ensure that the resonance frequency and the resonance linewidth $\Gamma_\mathrm{m}$ depend linearly on the laser power, we estimate $\Delta f_\mathrm{m}/\Delta P$ and $\Delta \Gamma_\mathrm{m}/\Delta P$ for absorbed laser powers below 35~nW when the temperature is below 40~K, and 60~nW otherwise. We emphasize that the temperature profile over the surface of the drum in the measurement of $\Delta f_\mathrm{m}/\Delta P$ differs from that of $\Delta f_\mathrm{m}/\Delta T$. This results in a prefactor in the conversion from the equivalent thermal conductance $K$ into the thermal conductivity of the monolayer, as described below.

\begin{figure*}
    \includegraphics[width=12cm]{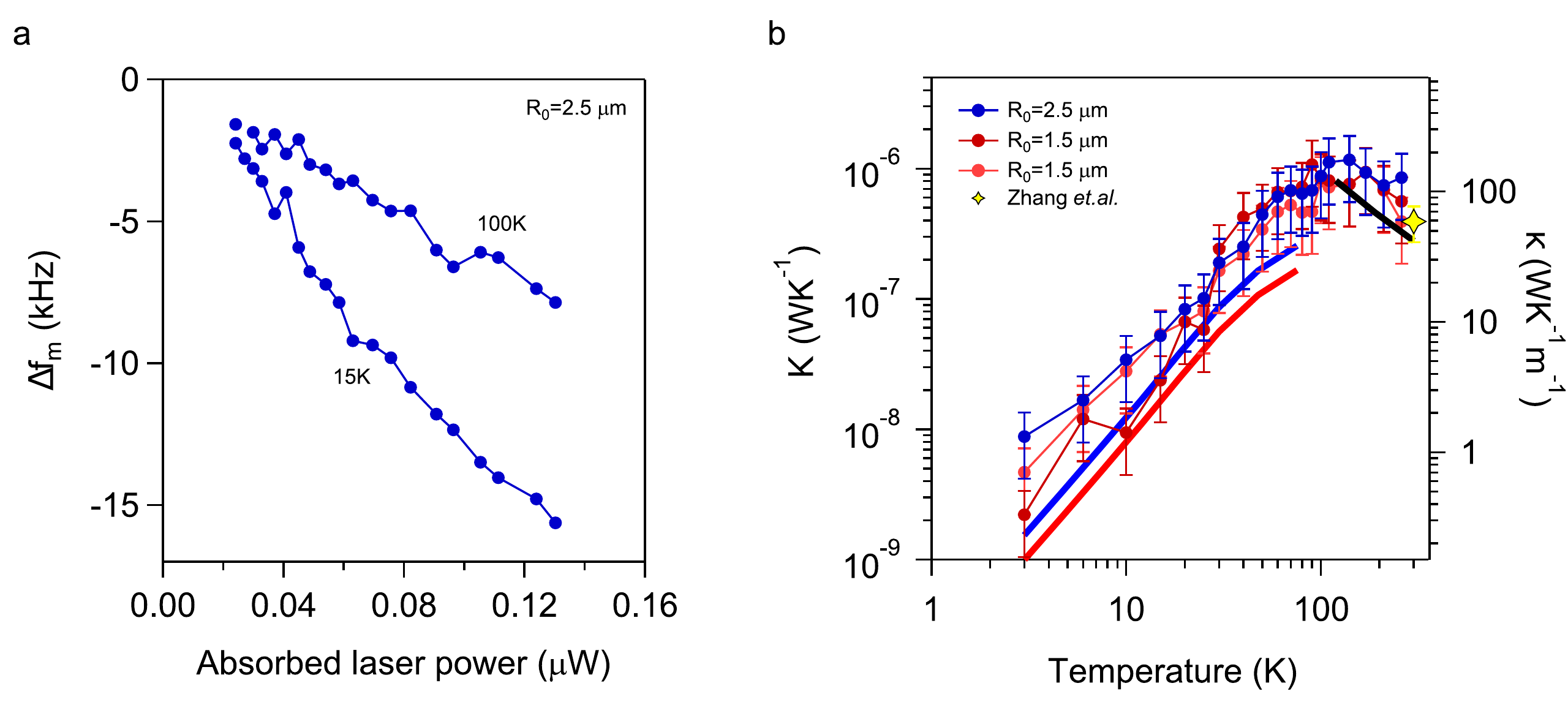}
	\caption{\textbf{Thermal conductance of MoSe$_2$ monolayers.} (\textbf{a}) Shift of the resonance frequency $\Delta f_\mathrm{m}$ as a function of absorbed laser power $P$ when the drum is in the straight configuration. (\textbf{b}) Thermal conductance $K=P/\Delta T$ as a function of temperature. The right axis shows the conductivity in the diffusive regime, which is obtained using Eq.~\ref{eq:conductivityconductance} with $\eta$=0.61. The yellow star symbol at 300~K corresponds to the thermal conductivity measured with the Raman method~\cite{Zhang2015}; we are not aware of another measurement of the thermal conductivity of MoSe$_2$ monolayers. The black line shows the conductivity in the diffusive regime for an infinitely large monolayer computed by solving the Boltzmann transport equation as in Ref.~\cite{Fugallo2013}. The red and the blue line corresponds to the conductance in the ballistic regime computed from first principles for the 1.5 and the 2.5~$\mu$m radius drum, respectively, using Eq.~\ref{eq:conductance} with $\alpha=2.1$ and $\alpha=3.2$.}
	\label{fig4}
\end{figure*}

The temperature dependance of the thermal conductance suggests diffusive transport at high temperature (Fig.~4b). Upon increasing temperature above $\sim 100$~K, the conductance decreases, which is attributed to the reduction of the mean-free path due to phonon-phonon scattering~\cite{Taube2015}. Below $\sim 100$~K, the conductance gets larger when increasing temperature. This indicates that phonon-phonon scattering is no more relevant, so that the mean-free path could be limited by e.g. the device size or the grain boundaries of the crystal. The error bars of the thermal conductance in Fig.~4b come from the uncertainty in the absorption coefficient $A$ of the monolayer (Section 2 of Supplementary Information). Since we cannot measure the absorption coefficient, we choose a large uncertainty, that is, $A=0.057 \pm 0.03$. Figure~4b shows that this affects the measured temperature dependence of the thermal conductance only weakly.

We measure $\tau$ by comparing the resonance frequency and the resonance linewidth measured with the resonator in the straight configuration ($V_\mathrm{g}^\mathrm{dc}=0$~V) and in the bending configuration ($V_\mathrm{g}^\mathrm{dc}=4$~V) (Section 4 of Supplementary Information). Here $V_\mathrm{g}^\mathrm{dc}=4$~V is the largest voltage that we apply, since a larger voltage may collapse the drum onto the bottom of the trench. The associated strain is less than 1\% (Section 3 of Supplementary Information). Such a small strain is expected to have no sizeable effect on the thermal transport properties~\cite{Fugallo2014}.

Figures 5(a-c) show that $\tau$ remains constant when varying temperature within the error bars of the measurements. We cannot measure $\tau$ above 100~K, since the reduced quality-factor prevents us to resolve $\Delta f_\mathrm{B}$. Using the averaged phonon velocity $v\simeq 1300$~m/s computed by first principles (Section 5 of Supplementary Information), the average time $\langle \tau \rangle =3.3 \pm 2.1 $~ns results in a mean-free path of about $4.3 \pm 2.7~\mu$m, which is consistent with the $2.5~\mu$m radius of the drum. This suggests that the majority of the phonon carriers are ballistic over the size of the drum.

\begin{figure*}
	\includegraphics[width=12cm]{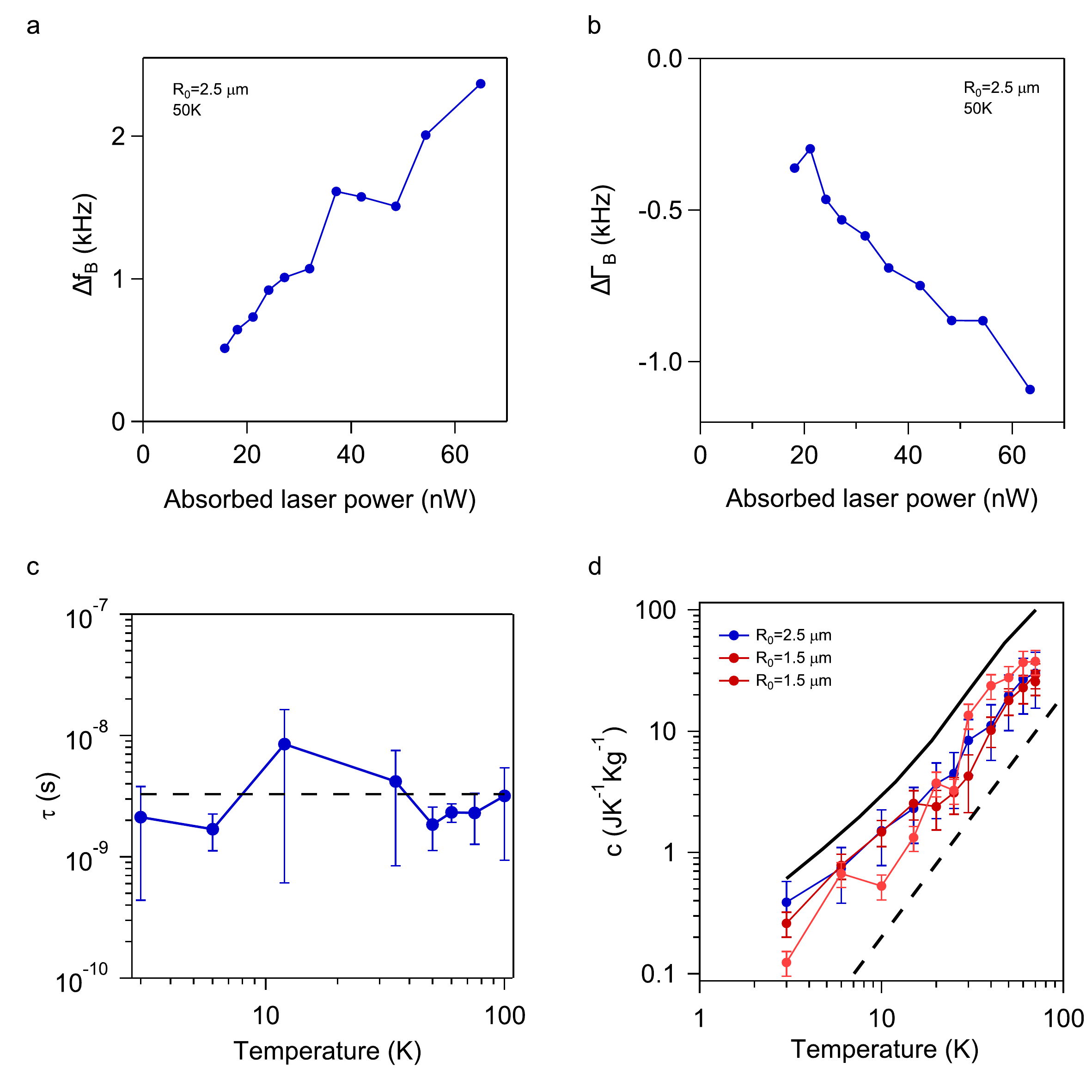}
	\caption{\textbf{Time for the heat to travel out of the drum and specific heat capacity of MoSe$_2$ monolayers.} (\textbf{a,b}) Shifts of the resonance frequency $\Delta f_\mathrm{B}$ and the mechanical bandwidth $\Delta \Gamma_\mathrm{B}$ as a function of absorbed laser power $P$. We obtain $\Delta f_\mathrm{B}$ and $\Delta \Gamma_\mathrm{B}$ by subtracting the frequency shift and the bandwidth shift measured in the bending configuration from that measured in the straight configuration. (\textbf{c}) Time for the heat to travel out of the drum as a function of temperature. The large error bars at 12 and 35~K are due to the drift of the resonance frequency caused by the automatized heating and cooling switches in our cryofree cryostat. The dashed black line corresponds to the averaged $\tau$. (\textbf{d}) Specific heat capacity as a function of temperature. We convert $C=\langle \tau \rangle K$ into $c$ using Eq.~\ref{eq:heatcapacity} with $\beta =0.86$. The black dashed line corresponds to the $T^\mathrm{2}$ dependence. The black continuous line corresponds to the specific heat capacity computed from first-principles. Since the displacement sensitivity of the 1.5~$\mu$m radius drums was not good enough to measure $\tau$, we estimate $\langle \tau \rangle$ from the value measured with the 2.5 $\mu$m radius drum and the radius ratio. The error bars come from the uncertainty in $\langle \tau \rangle$ and $K$.}
	\label{fig5}
\end{figure*}

These measurements allow us to directly quantify the equivalent heat capacity of an individual MoSe$_2$ monolayer using $C=\langle \tau \rangle K$. Figure~5d shows that the temperature dependence of the heat capacity approaches a $T^2$ dependence. This is consistent with the $T^d$ dependence expected for two-dimensional systems in its simplest form, where $d=2$ is the dimensionality. Previous measurements of the phononic heat capacity of nano-materials were carried out by packing them in macroscopic ensembles, such as films of nanotube ropes~\cite{Hone2000a} and powders of MoSe$_2$ multilayered crystals~\cite{Kiwia1975}. Such ensemble measurements suffer from the coupling between nano-systems, which modifies the heat capacity at low temperature.

The temperature profile along the heat flow has to be considered when evaluating the specific heat capacity $c$ and the thermal conductivity $\kappa$ of MoSe$_2$ monolayers (Figs.~4b and 5d). The temperature is non-uniform over the surface of the drum when measuring the slope $\Delta f_\mathrm{m}/\Delta P$, while it is uniform during the measurement of the calibration slope $\Delta f_\mathrm{m}/\Delta T$. These different temperature profiles add a geometrical constant in the conversion from $C$ and $K$ into $c$ and $\kappa$. In the ballistic regime, the temperature is taken as constant within a disc corresponding to the region illuminated by the laser beam of radius $r_0$; outside this region, the temperature drops as $1/r$ along the radial coordinate $r$ because of the conservation of heat flow in our disc geometry (Section 5 of Supplementary Information). This contrasts with the constant temperature profile along ballistic conductors with uniform width. In the diffusive regime, the temperature decreases logarithmically along $r$ due to phonon scattering events~\cite{Cai2010,Johnson2013,Cepellotti2017}. The measured $C$ and $K$ are converted into $c$ and $\kappa$ using
\begin{eqnarray}
&& c=\frac{C}{\pi R_0^2t\rho } \beta,
\label{eq:heatcapacity}\\
&& \kappa= \frac{K}{2\pi t} \eta,
\label{eq:conductivityconductance}
\end{eqnarray}
where $R_0$ is the radius of the suspended drum, $t=0.64$~nm the thickness of the monolayer, and $\rho$ the mass density of MoSe$_2$. The geometrical constants $\beta$ and $\eta$ are of the order of one and depend on $R_0$, $r_0$, and the temperature profile (Section 5 of Supplementary Information). The conductivity in Fig.~4b is determined in the diffusive regime only.

The measured temperature dependence of $\kappa$ above $\sim 100$~K can be described by first-principles calculations on MoSe$_2$ monolayers in the diffusive regime (Fig.~4b), whereas the measured temperature dependencies of $c$ and $K$ below $\sim 100$~K are consistent with predictions in the ballistic regime (Figs.~4b and 5d). For the comparison between measurements and theory, we derive the ballistic conductance in our peculiar disc geometry assuming that the inner reservoir is given by the radius $r_0$, and the outer reservoir by $R_0$. We obtain
\begin{eqnarray}
&& K= 2\pi r_0t\alpha \cdot \frac{\rho c v}{2} ,
\label{eq:conductance}\\
&& v=\frac{\sum_\mathrm{q,s} C_\mathrm{q,s} \frac{2|v_\mathrm{q,s}|}{\pi}}{\sum_\mathrm{q,s}C_\mathrm{q,s}},
\label{eq:velocity}
\end{eqnarray}
where $C_\mathrm{q,s}=\frac{dn_\mathrm{q,s}}{dT}\hbar \omega_\mathrm{q,s}$ is the specific heat of the phonon of the branch $s$ with momenta $q$, $\omega_\mathrm{q,s}$ the phonon pulsation, $n_\mathrm{q,s}$ the Bose occupation factor, and $v_\mathrm{q,s}$ the group velocity. The constant $\alpha$ is another geometric factor of the order of one like $\beta$ and $\eta$. The expression of these three geometrical factors is given in Eqs. S34, S50, and after S26 of Supplementary Information. The phonon properties of the monolayer lattice are calculated using density functional perturbation theory. Instead, in the diffusive regime, the conductivity is derived by an exact solution of the Boltzmann transport equation taking into account three-phonon interactions and isotopic scattering~\cite{Fugallo2013}. In such a calculation, we use scattering rates derived by first principles that depend on the energy and momentum of the involved phonons, in contrast to the single empirical effective time $\tau$ used in Eqs.~\ref{DfB} and \ref{GFB}, which describes the characteristic time for the heat to travel out of the resonator. The conductivity derived with a homogenous temperature gradient ($\nabla T$) can be compared to the measured conductance through Eq.~\ref{eq:conductivityconductance}, which maps transport with non-homogenous $\nabla T$ to that with homogenous $\nabla T$. The derivation of Eqs.~\ref{eq:heatcapacity}-\ref{eq:velocity} and information on the first-principle calculations can be found in Section 5 of Supplementary Information. The reasonably good agreement between measurement and theory in Fig.~4b suggests that the resistance at the interface between the monolayer and the substrate does not contribute significantly to the thermal transport. Future work will be carried out on smaller diameter drums where the resistance of the interface is expected to become comparatively larger.

Our optomechanical measurements provide a detailed picture of thermal transport in monolayer MoSe$_2$ lattices down to cryogenic temperature. Our work opens the possibility to measure thermal properties in a large variety of different two-dimensional materials, because the devices required for these measurements are simple to fabricate. We will improve the quality factor of drums by e.g. increasing their diameter in order to measure $\tau$ and the heat capacity up to room temperature. This new measurement method may allow the exploration of the phonon hydrodynamics regime, which is expected to be robust in monolayer systems~\cite{Cepellotti2015,Lee2015b}. This regime is interesting because heat is carried by collective excitations of phonon states. This gives rise to a new type of sound propagation, called second sound. The measurement of $\tau$ should enable the direct access of the velocity of the second sound. In addition, this new measurement method may shed light on the divergence of the thermal conductivity in two-dimensions, when the size of the system increases~\cite{Xu2014}. The origin of this behaviour is under active investigation with different interpretations based on either the dimensionality of the system or the special phononic states that remain ballistic over extraordinarily long distances~\cite{Lepri2003,Lindsay2014,Fugallo2014}. Optomechanical measurements also enable the study of the anisotropic thermal conductivity, as recently demonstrated in $10-100$~nm thick black phosphorus crystals at room temperature~\cite{Islam2018}.

\textbf{Acknowledgments} This work is supported by the ERC advanced grant 692876, the Foundation Cellex, the CERCA Programme, AGAUR, Severo Ochoa (SEV-2015-0522), the grant FIS2015-69831-P of MINECO, the Fondo Europeo de Desarrollo Regional (FEDER), and the Project P2ELP2-168546 of the Swiss National Science Foundation.

{\textbf{Author contributions}
NM fabricated the devices. NM, ST, and ARP carried out the experiment with support from MM and XM. AC did the first-principles calculations. AC and FM developed the model of the thermal transport in the ballistic regime with contributions from AI and AB. IE carried out the simulations of the interference pattern of the laser beam. The data analysis was done by NM and AB. NM and AB wrote the manuscript with comments from the other authors. AB supervised the work.}

\bibliographystyle{apsrev4-1}

\end{document}